\documentclass[conference]{IEEEtran}
\usepackage[left=0.67in,right=0.67in,top=0.75in,bottom=0.80in]{geometry}
\usepackage[T1]{fontenc}
\usepackage[latin9]{inputenc}
\usepackage{amsthm}
\usepackage{amsmath}
\usepackage{bbm} 
\usepackage{graphicx} 
\usepackage{color} 
\usepackage{cite}
\usepackage{algpseudocode}
\usepackage{algorithm}
\usepackage{amssymb}
\usepackage{subfigure}
\usepackage{esint}
\usepackage{algpseudocode}
\usepackage{epstopdf}
\usepackage{multirow}
\usepackage{makecell}
\usepackage{float}
\usepackage{lipsum}
\usepackage{balance}

\newtheorem{definition}{Definition}
\newtheorem{remark}{Remark}

\newtheorem{example}{Example}

\theoremstyle{plain}
\theoremstyle{plain}
\newtheorem{theorem}{Theorem}
\newtheorem{lemma}{Lemma}

\newcommand{\comment}[1]{}

\IEEEoverridecommandlockouts

\begin{document}

\title{Minimizing the Number of Detrimental Objects in Multi-Dimensional Graph-Based Codes}

\author{
   \IEEEauthorblockN{Ahmed Hareedy, Rohith Kuditipudi, and Robert Calderbank}
   \IEEEauthorblockA{Electrical and Computer Engineering Department, Duke University, Durham, NC 27705 USA \\ ahmed.hareedy@duke.edu, rohith.kuditipudi@duke.edu, and robert.calderbank@duke.edu}
}
\maketitle

\begin{abstract}
In order to meet the demands of data-hungry applications, data storage devices are required to be increasingly denser. Various sources of error appear with this increase in density. Multi-dimensional (MD) graph-based codes are capable of mitigating error sources like interference and channel non-uniformity in dense storage devices. Recently, a technique was proposed to enhance the performance of MD spatially-coupled codes that are based on circulants. The technique carefully relocates circulants to minimize the number of short cycles. However, cycles become more detrimental when they combine together to form more advanced objects, e.g., absorbing sets, including low-weight codewords. In this paper, we show how MD relocations can be exploited to minimize the number of detrimental objects in the graph of an MD code. Moreover, we demonstrate the savings in the number of relocation arrangements earned by focusing on objects rather than cycles. Our technique is applicable to a wide variety of one-dimensional (OD) codes. Simulation results reveal significant lifetime gains in practical Flash systems achieved by MD codes designed using our technique compared with OD codes having similar parameters.
\end{abstract}

\section{Introduction}\label{sec_intro}

The continuous and rapid growth in the density of modern storage devices brings many challenges. One of these challenges is an increase in the number of sources of data corruption in the system, which requires advanced error correcting codes to be applied. Because of their capacity-approaching performance and the degrees of freedom they offer in the code construction, graph-based codes, e.g., low-density parity-check (LDPC) codes, are applied in many data storage systems. Binary and non-binary graph-based codes are used in both Flash \cite{mit_nl, ahh_tit} and magnetic recording \cite{shafa} systems to significantly improve the performance.

Multi-dimensional (MD) graph-based codes are constructed by coupling different copies of a one-dimensional (OD) code to enhance the code properties. Because of the additional design flexibility offered by MD coupling, MD codes are capable of alleviating different types of interference and channel non-uniformity in modern storage systems. One example is mitigating inter-track interference in two-dimensional magnetic recording (TDMR) systems \cite{shafa} through specific non-binary LDPC code constructions as in \cite{chen_2d}. Various MD spatially-coupled (MD-SC) codes have been presented in the literature \cite{truhach, ohashi, schmalen, liu}. While these MD-SC codes demonstrated performance gains, they had limitations in the underlying OD codes and the topologies of the resulting MD codes.

Recently, the authors of \cite{md_res} proposed a technique for a systematic construction of MD-SC codes that are based on circulants. Through carefully chosen relocations of circulants from the copies of the OD code to certain auxiliary matrices, they managed to significantly reduce the number of short cycles in the graph of the MD-SC code. While cycles are not preferred in graph-based codes, they become a lot more detrimental when they combine together to form absorbing sets (ASs), including low-weight codewords. ASs, not cycles, are the objects that dominate the error profile of graph-based codes in the error floor region \cite{ahh_tit, lara_as}.

In this paper, we demonstrate how to use MD coupling to eliminate as many detrimental objects as possible from the graph of an MD code. The underlying OD codes we use can be structured or random, and can be block or SC codes. By deriving the fraction of relocation arrangements for different cases, we manifest the savings in relocation options achieved by operating on objects rather than cycles. Experimental results emphasizing the reduction in the multiplicity of detrimental objects are shown. Simulation results demonstrating $\approx 1200$ (resp., $1800$) program/erase cycles gain in the waterfall (resp., error floor) region over practical Flash channels compared with OD codes having similar length and rate are presented.

The rest of the paper is organized as follows. In Section~\ref{sec_prelim}, MD graph-based codes are introduced. How ASs are removed via relocations is discussed in Section~\ref{sec_rem}. Next, the savings in relocation arrangements are derived in Section~\ref{sec_save}. In Section~\ref{sec_exp}, the code design algorithm and experimental results are presented. The paper is concluded in Section~\ref{sec_conc}.

\section{MD Graph-Based Codes}\label{sec_prelim}

The technique we propose in this paper can be used to construct binary and non-binary codes. However, since the process of relocations affects only the code topology, we focus on the unlabeled graphs (all edge weights are set to $1$) and binary matrices \cite{ahh_tit}.

Define $\bold{H}_{\textup{OD}}$ as the parity-check matrix of the underlying OD code, and $\bold{H}_{\textup{MD}}$ as the parity-check matrix of the MD code. Recall the correspondence between the parity-check matrix and the graph of a code. Define $M-1$ auxiliary matrices, $\bold{X}_1$, $\bold{X}_2$, \dots, $\bold{X}_{M-1}$ having the same dimensions as $\bold{H}_{\textup{OD}}$. The MD matrix $\bold{H}_{\textup{MD}}$ is given by:
\begin{gather}\label{eqn_hmd}
\bold{H}_{\textup{MD}} \triangleq
\begin{bmatrix}
\bold{H}'_{\textup{OD}} & \bold{X}_{M-1} & \bold{X}_{M-2} & \dots & \bold{X}_2 & \bold{X}_1 \vspace{-0.0em}\\
\bold{X}_1 & \bold{H}'_{\textup{OD}} & \bold{X}_{M-1} & \dots & \bold{X}_3 & \bold{X}_2 \vspace{-0.0em}\\
\bold{X}_2 & \bold{X}_1 & \bold{H}'_{\textup{OD}} & \dots & \bold{X}_4 & \bold{X}_3 \vspace{-0.0em}\\
\vdots & \vdots & \vdots & \vdots & \vdots & \vdots \vspace{-0.0em}\\
\bold{X}_{M-2} & \bold{X}_{M-3} & \bold{X}_{M-4} & \dots & \bold{H}'_{\textup{OD}} & \bold{X}_{M-1} \vspace{-0.0em}\\
\bold{X}_{M-1} & \bold{X}_{M-2} & \bold{X}_{M-3} & \dots & \bold{X}_1 & \bold{H}'_{\textup{OD}}
\end{bmatrix},
\end{gather}
where $\bold{H}_{\textup{OD}} = \bold{H}'_{\textup{OD}} + \sum_{\ell=1}^{M-1} \bold{X}_\ell$ (see also \cite{md_res}). The graphs of the OD codes we use do not have any cycles of length $4$. According to the construction of $\bold{H}_{\textup{MD}}$, the data to be stored is separated into $M$ chunks, and each chunk is stored in a track or a sector of the storage device. The matrix $\bold{H}_{\textup{MD}}$ is constructed by coupling the $M$ OD copies of $\bold{H}_{\textup{OD}}$~via carefully relocating some of the non-zero (NZ) entries in these copies to auxiliary matrices in order to eliminate certain detrimental objects from the graph of the MD code. Relocations are mathematically represented by an MD mapping as follows:
\begin{equation}\label{eq_reloc}
R: \{\mathcal{E}_{i,j}, \forall i, j\} \rightarrow \{0, 1, \dots, M-1\},
\end{equation}
where $\mathcal{E}_{i,j}$ is an NZ entry corresponding to an edge connecting check node (CN) $i$ to variable node (VN) $j$ in the graph of $\bold{H}_{\textup{OD}}$. This mapping is explained as follows: $R \left (\mathcal{E}_{i,j} \right ) = \ell > 0$ means that the NZ entry $\mathcal{E}_{i,j}$ is relocated from $\bold{H}_{\textup{OD}}$ to $\bold{X}_\ell$ ($M$ times) at the same position $(i, j)$ it had in $\bold{H}_{\textup{OD}}$, with $R \left (\mathcal{E}_{i,j} \right ) =0$ referring to the no-relocation case.

Here, $M$ is a prime integer $> 2$, and both $\bold{H}_{\textup{OD}}$ and $\bold{H}_{\textup{MD}}$ have a fixed column weight, i.e., fixed VN degree, $\gamma$. The row weight, i.e., CN degree, is not necessarily fixed.

Define a cycle of length $2k$ in the graph of $\bold{H}_{\textup{OD}}$ by the following set of NZ entries in $\bold{H}_{\textup{OD}}$: $\{\mathcal{E}_{i_1,j_1}, \mathcal{E}_{i_2,j_2}, \dots \mathcal{E}_{i_{2k},j_{2k}} \}$, such that two entries $\mathcal{E}_{i_w,j_w}$ and $\mathcal{E}_{i_{w+1},j_{w+1}}$, $1 \leq w \leq 2k$ and $\mathcal{E}_{i_{2k+1},j_{2k+1}} = \mathcal{E}_{i_1,j_1}$, are consecutive entries on the cycle. The authors of \cite{md_res} proved that this cycle stays \textbf{active} after a relocation arrangement if and only if\footnote{This condition bares similarity to the condition in \cite{fos_lift} for protograph lifting. In fact, some of the results in this paper are applicable to the procedures of lifting and non-binary labeling.}:
\begin{equation}\label{eqn_cyc}
\sum_{w=1}^{2k} (-1)^w R(\mathcal{E}_{i_w,j_w}) \equiv 0 \textup{ (mod $M$)}.
\end{equation}
For a cycle of length $2k$ to stay active, its $M$ copies must~result in $M$ cycles of length $2k$ in the graph of $\bold{H}_{\textup{MD}}$. If (\ref{eqn_cyc}) is not satisfied, the cycle becomes \textbf{inactive}, and its $M$ copies result in a single cycle of length $2kM$. The result in \cite{md_res} was for $M=3$. However, this result generalizes to any prime $M$.

Under iterative decoding, the detrimental (error-prone) objects in the graph of a code are typically ASs, including low-weight codewords. This was shown to be the case for additive white Gaussian noise (AWGN) \cite{lara_as, behzad_elem}, Flash \cite{ahh_tit, ahh_nboo}, and magnetic recording \cite{ahh_tit} channels. Thus, recall:

\begin{definition}\label{uas}
Let $\mathcal{V}$ be a subset of VNs in the unlabeled graph of a code. Let $\mathcal{O}$ (resp., $\mathcal{T}$ and $\mathcal{H}$) be the set of degree-$1$ (resp., $2$ and $> 2$) CNs connected to $\mathcal{V}$. This graphical configuration is an $(a, d_1)$ unlabeled elementary absorbing set (UAS) if $|\mathcal{V}| = a$, $\vert{\mathcal{O}}\vert=d_1$, $\vert{\mathcal{H}}\vert=0$, and each VN in $\mathcal{V}$ is connected to strictly more neighbors in $\mathcal{T}$ than in $\mathcal{O}$.
\end{definition}

\begin{remark}\label{rmk_nelem}
Many non-elementary absorbing sets appearing in the error profile of non-binary graph-based codes over practical Flash channels have underlying unlabeled elementary configurations \cite{ahh_tit}.
\end{remark}

\begin{figure}
\vspace{-0.9em}
\center
\includegraphics[trim={0.0in 0.5in 0.0in 0.1in}, width=3.5in]{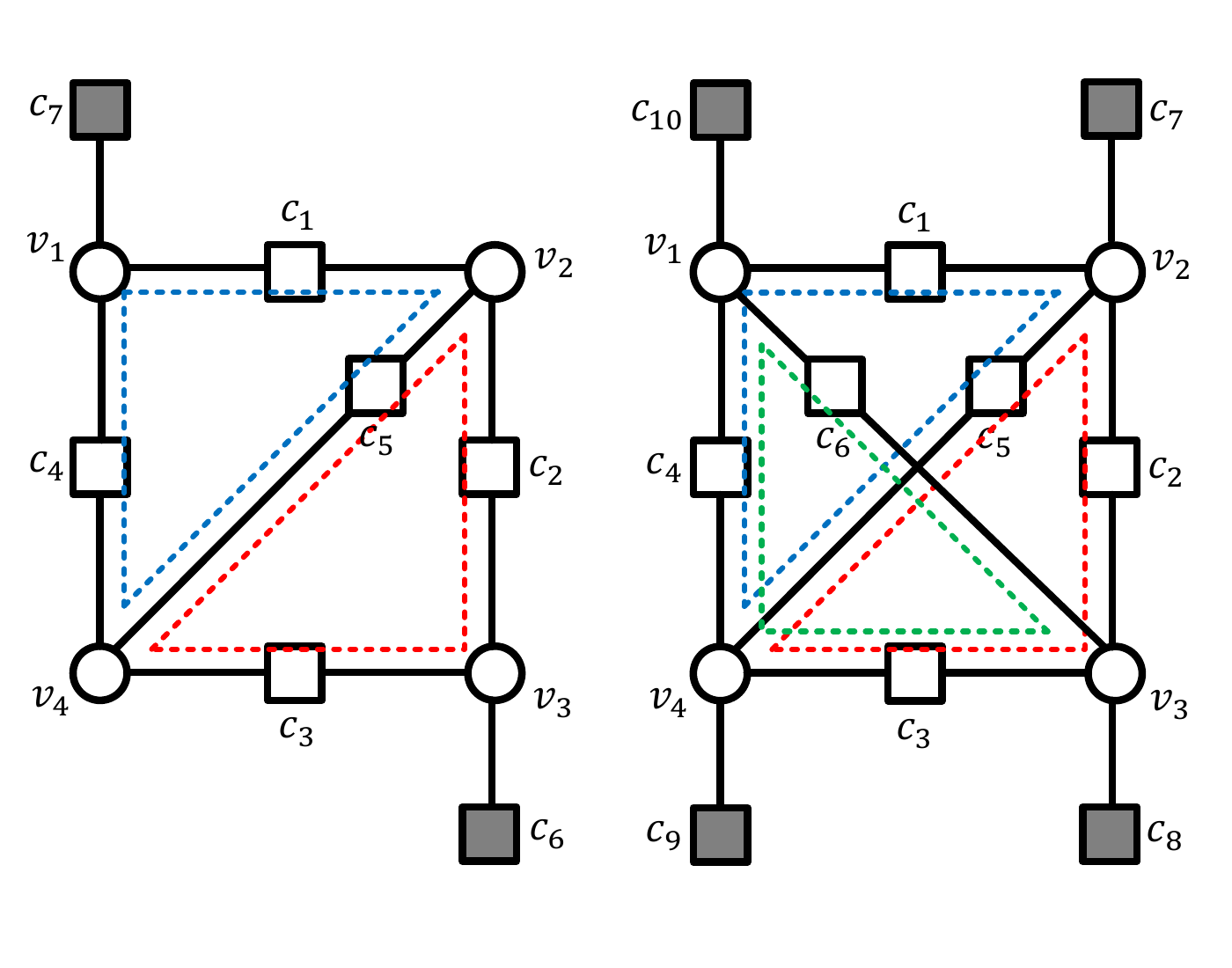}
\vspace{-0.9em}
\caption{Left panel: a $(4, 2)$ UAS, $\gamma=3$. Right panel: a $(4, 4)$ UAS, $\gamma=4$. Basic cycles are shown in dotted lines.}
\label{fig_1}
\vspace{-0.5em}
\end{figure}

We study UASs having connected subgraphs. A $(4, 2)$ UAS in a code with $\gamma=3$ and a $(4, 4)$ UAS in a code with $\gamma=4$ are shown in Fig.~\ref{fig_1}. Circles (resp., grey squares and white squares) represent VNs (resp., degree-$1$ CNs and degree-$2$ CNs). In the following sections, we will investigate how to perform relocations to minimize the number of UASs in the graph of an MD code to enhance its performance.

\section{Removing ASs Through Relocations}\label{sec_rem}

An $(a, d_1)$ UAS has the following number of degree-$2$ CNs:
\begin{equation}\label{eqn_d2}
d_2 = \frac{1}{2} (a\gamma - d_1).
\end{equation}
We now revisit the concept of \textbf{basic cycles}, which generalizes the concept of fundamental cycles first introduced in \cite{behzad_elem} for non-binary graph-based codes, to represent a UAS.

\begin{definition}\label{fund_cycs}
A cycle basis $\mathcal{B}_\textup{c}$ of an $(a, d_1)$ UAS is a minimum-cardinality set of cycles using disjunctive unions of which, each~cycle in the UAS can be obtained. We call the cycles in $\mathcal{B}_\textup{c}$ basic cycles.
\end{definition}

Denote a Galois field of size $q$ as GF($q$). Since our graphs are unlabeled (no weights), span$(\mathcal{B}_{\textup{c}})$ can be represented by a vector space over GF($2$), with its vectors being of size $n_\textup{e}=2d_2$ and their elements are also in $\textup{GF}($2$)$. There are $n_\textup{f} = \vert \mathcal{B}_{\textup{c}} \vert$ basic cycles. From graph theory principles, this number is computed by subtracting the number of degree-$2$ CNs, each represented by the pair of edges adjacent to it, comprising the tree spanning all VNs from the total number of degree-$2$ CNs. Consequently,
\begin{align}\label{eqn_nf}
n_\textup{f} &= \frac{1}{2} \left (n_\textup{e} - 2(a-1) \right ) = d_2 - a + 1 \nonumber \\ &= \frac{1}{2} \left ( a(\gamma-2) - d_1 + 2 \right ),
\end{align}
where the last equality is obtained using (\ref{eqn_d2}). Without loss of generality, in this paper, we always select the basic cycles in $\mathcal{B}_{\textup{c}}$ to be of the smallest lengths for simplicity.

\begin{example}\label{ex_1}
Consider the $(4, 2)$ UAS, $\gamma=3$, in Fig.~\ref{fig_1}. From (\ref{eqn_nf}), the number of basic cycles is:
\begin{equation}
n_\textup{f} = \frac{1}{2} \left ( 4(3-2) - 2 + 2 \right ) = 2. \nonumber
\end{equation}
We select the two cycles in dotted blue and dotted red shown in the figure to be the elements of $\mathcal{B}_\textup{c}$. A cycle in span$(\mathcal{B}_{\textup{c}})$ can be written as:
\vspace{-0.1em}\begin{align}
\big [ &e_{c_1,v_1} \text{ } e_{c_1,v_2} \text{ } e_{c_2,v_2} \text{ } e_{c_2,v_3} \text{ } e_{c_3,v_3} \text{ } e_{c_3,v_4} \text{ } \nonumber \\ &e_{c_4,v_4} \text{ } e_{c_4,v_1} \text{ } e_{c_5,v_2} \text{ } e_{c_5,v_4} \big ], \nonumber
\end{align}
where $e_{i_w,j_w} = \mathbbm{1} \left ( \mathcal{E}_{i_w,j_w} \right )$ is an indicator function of the existence of the NZ entry $\mathcal{E}_{i_w,j_w}$. Thus, the dotted blue and dotted red basic cycles are:
\begin{equation}
[ 1 \text{ } 1 \text{ } 0 \text{ } 0 \text{ } 0 \text{ } 0 \text{ } 1 \text{ } 1 \text{ } 1 \text{ } 1 ] \text{ and } [ 0 \text{ } 0 \text{ } 1 \text{ } 1 \text{ } 1 \text{ } 1 \text{ } 0 \text{ } 0 \text{ } 1 \text{ } 1 ], \nonumber
\end{equation}
respectively. Adding the vectors of the two basic cycles over GF($2$) gives:
\begin{equation}
[ 1 \text{ } 1 \text{ } 1 \text{ } 1 \text{ } 1 \text{ } 1 \text{ } 1 \text{ } 1 \text{ } 0 \text{ } 0 ], \nonumber
\end{equation}
which is the vector of the remaining cycle in the UAS.
\end{example}

Given the number of basic cycles in an $(a, d_1)$ UAS, we now introduce useful bounds on the total number of cycles.

\begin{lemma}\label{lem_ncyc}
The total number of cycles, $n_{\textup{c}}$, in an $(a, d_1)$ UAS having $n_{\textup{f}}$ basic cycles is bounded as follows:
\begin{equation}\label{eqn_ncb}
\frac{1}{2}n_{\textup{f}}(n_{\textup{f}}+1) \leq n_{\textup{c}} \leq 2^{n_{\textup{f}}}-1.
\end{equation}
\end{lemma}

\begin{IEEEproof}
\textbf{Lower bound:} Since the subgraph of the UAS is connected, we can always find an order for the basic cycles such that each two consecutive basic cycles share at least one degree-$2$ CN. Any two cycles sharing at least one degree-$2$ CN form a new cycle if the vectors representing them are added. Thus, the minimum value of $n_{\textup{c}}$ is computed as follows. At first, we have one cycle, which is the first basic cycle. Then, we get two more cycles, which are the second basic cycle and the cycle resulting from adding the vectors of the first and the second ones over GF($2$); we refer to this cycle as cycle $(1, 2)$. Then, we get at least three more cycles referred to as $3$, $(2, 3)$, and $(1, 2, 3)$, and the lower bound is achieved if the first and third basic cycles do not share CNs. This continues till the last basic cycle. As a result,
\begin{equation}\label{eqn_lower}
n_{\textup{c}} \geq \sum_{\delta=1}^{n_{\textup{f}}} \delta = \frac{1}{2}n_{\textup{f}}(n_{\textup{f}}+1).
\end{equation}

\textbf{Upper bound:} The upper bound is achieved if the addition of any distinct group of basic cycles gives a distinct cycle. Consequently,
\begin{equation}\label{eqn_upper}
n_{\textup{c}} \leq \sum_{\delta=1}^{n_{\textup{f}}} \binom{n_{\textup{f}}}{\delta} = 2^{n_{\textup{f}}}-1,
\end{equation}
where the second equality follows from the binomial theorem. Combining (\ref{eqn_lower}) and (\ref{eqn_upper}) gives (\ref{eqn_ncb}).
\end{IEEEproof}

\begin{example}\label{ex_2}
The upper and the lower bounds are the same for the $(4, 2)$ UAS, $\gamma=3$, in Fig.~\ref{fig_1}. Since $n_{\textup{f}}=2$, $n_{\textup{c}}=\frac{1}{2}2(2+1)=3$ from (\ref{eqn_ncb}), which is what we know from Example~\ref{ex_1}. On the contrary, only the upper bound is achieved for the $(4, 4)$ UAS, $\gamma=4$, in Fig.~\ref{fig_1} because of its connectivity. Since $n_{\textup{f}}=3$ from (\ref{eqn_nf}), $n_{\textup{c}}=2^3-1=7$ from (\ref{eqn_ncb}).
\end{example}

We are now ready to introduce the condition under which a UAS stays \textbf{active} after a relocation arrangement. For an $(a, d_1)$ UAS to stay active, its $M$ copies in the graphs of $\bold{H}_{\textup{OD}}$ copies must result in $M$ $(a, d_1)$ UASs in the graph of $\bold{H}_{\textup{MD}}$.

\begin{theorem}\label{thm_cond}
The necessary and sufficient condition for an $(a, d_1)$ UAS to stay active after a relocation arrangement is that (\ref{eqn_cyc}) is satisfied for all the $n_{\textup{f}}$ basic cycles in a cycle basis $\mathcal{B}_{\textup{c}}$ of the UAS. Otherwise, the UAS becomes inactive, and the $Ma$ VNs of its $M$ copies form an $(Ma, Md_1)$ object.
\end{theorem}

\begin{IEEEproof}
We prove the sufficiency of the condition in~Theorem~\ref{thm_cond} first. The $(a, d_1)$ UAS stays active if all its $n_{\textup{c}}$ cycles stay active after the relocation arrangement, i.e., if (\ref{eqn_cyc}) is satisfied for all its cycles. By definition, any cycle in the UAS is a disjunctive union of the basic cycles, i.e., a linear combination of the vectors of the basic cycles, of that UAS. Thus, if (\ref{eqn_cyc}) is satisfied for all the $n_{\textup{f}}$ basic cycles, it is also satisfied for all the $n_{\textup{c}}$ cycles. Therefore, the UAS stays active if (\ref{eqn_cyc}) is satisfied for all its basic cycles in $\mathcal{B}_{\textup{c}}$.

The necessity follows from that if at least one basic cycle does not have (\ref{eqn_cyc}) satisfied after relocations, then there exists at least one cycle in the UAS that is not active. Thus, the UAS becomes inactive.

Now, if the UAS is inactive, at least one of its cycles has:
\begin{equation}\label{eqn_pr_11}
\sum_{w=1}^{2a'} (-1)^w R(\mathcal{E}_{i_w,j_w}) \not\equiv 0 \textup{ (mod $M$)},
\end{equation}
where $a' \leq a$ is the number of VNs in that cycle. Since we use prime $M$, the left-hand side becomes $0 \textup{ (mod $M$)}$ only via:
\begin{align}\label{eqn_pr_12}
M\sum_{w=1}^{2a'} (-1)^w R(\mathcal{E}_{i_w,j_w}) &\equiv 0 \textup{ (mod $M$)}, \text{ i.e., } \nonumber \\
\sum_{w=1}^{2Ma'} (-1)^w R(\mathcal{E}_{i_w,j_w}) &\equiv 0 \textup{ (mod $M$)},
\end{align}
which corresponds to a cycle of length $2Ma'$. This observation means that $Ma'$ VNs from the $M$ copies of the UAS form a cycle together after relocations. Consequently, and since the subgraph of the $(a, d_1)$ UAS is connected, the $Ma$ VNs of the $M$ copies of the UAS form an $(Ma, Md_1)$ object.
\end{IEEEproof}

If the UAS becomes inactive after relocations, its $M$ copies are \textbf{removed} from the graph of $\bold{H}_{\textup{MD}}$. Depending on certain factors, including which cycles in the $(a, d_1)$ UAS become inactive after relocations, different, possibly non-isomorphic, $(Ma, Md_1)$ configurations can be generated if the UAS is inactive. On a smaller scale, the M copies of the $(a, d_1)$ UAS result in multiple $(a, d_1+2\beta)$ objects, $\beta > 0$, in this case.

\begin{figure}
\vspace{-0.5em}
\center
\includegraphics[trim={0.5in 0.7in 0.5in 0.8in}, width=3.3in]{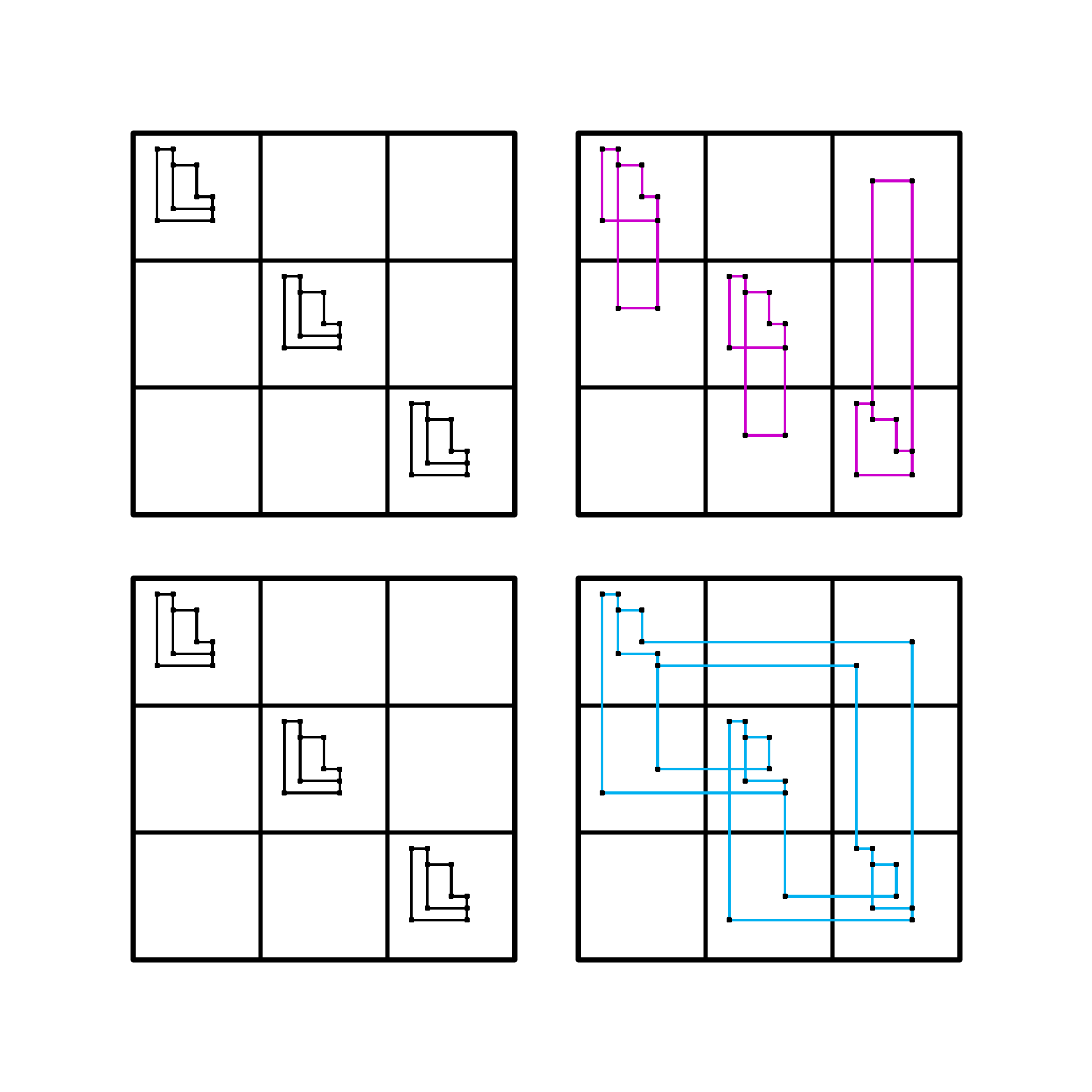}
\vspace{-0.9em}
\caption{Upper panel: Arrangement 1 is keeping three instances of the $(4, 2)$ UAS in $\bold{H}_{\textup{MD}}$. Lower panel: Arrangement 2 is removing the three copies of the $(4, 2)$ UAS from $\bold{H}_{\textup{MD}}$. VNs of the $(4, 2)$ UAS, which are columns in the matrix, are ordered from left to right as $v_1$, $v_2$, $v_3$, and $v_4$ (see Fig.~\ref{fig_1}).}
\label{fig_2}
\vspace{-0.3em}
\end{figure}

\begin{figure}
\vspace{-0.7em}
\center
\includegraphics[trim={0.8in 1.3in 0.9in 0.7in}, width=3.45in]{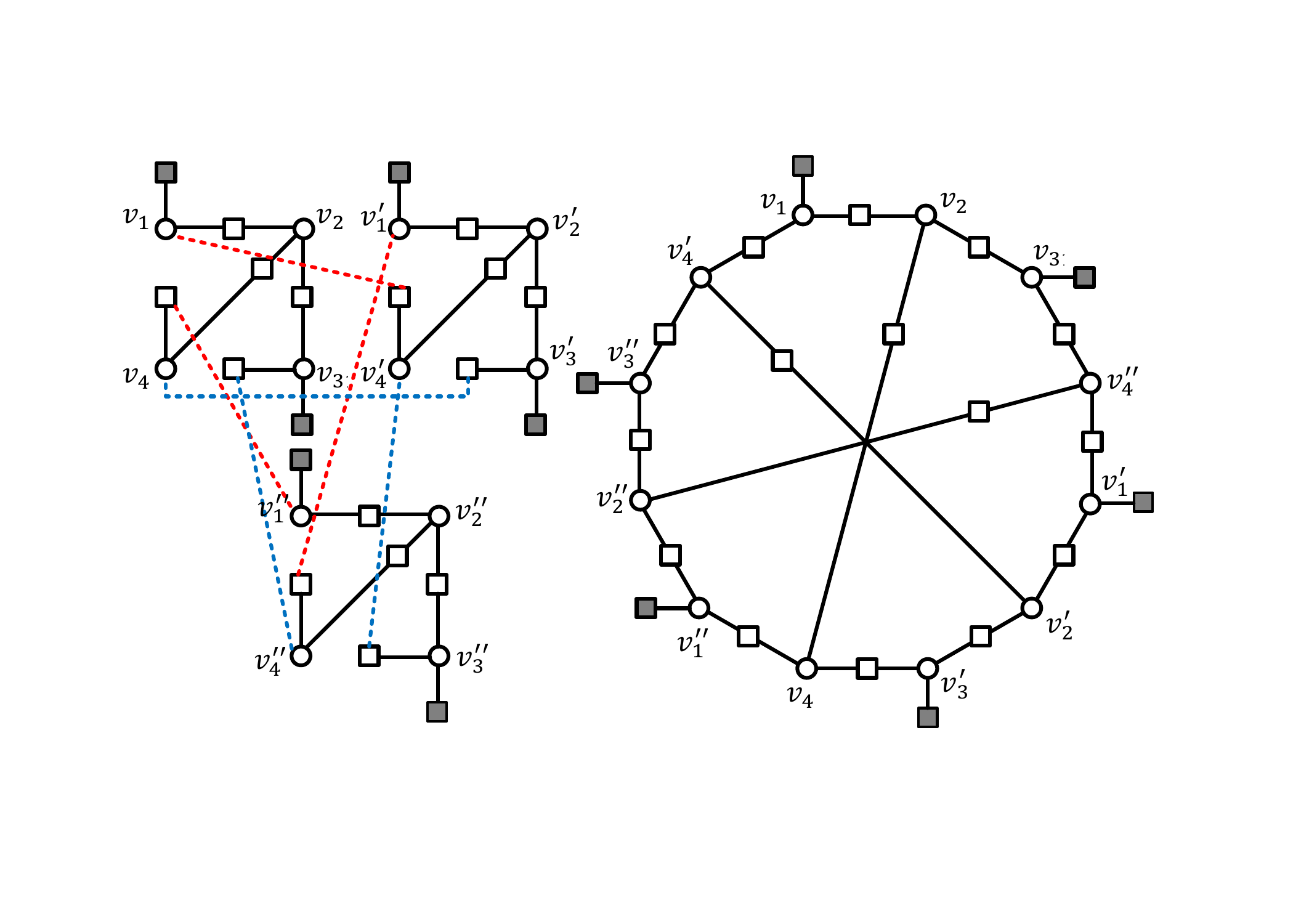}
\vspace{-0.3em}
\caption{Forming a $(12, 6)$ object from three copies of the $(4, 2)$ UAS after relocations. VNs in $\{v_1, v_2, v_3, v_4\}$ (resp., $\{v'_1, v'_2, v'_3, v'_4\}$ and $\{v''_1, v''_2, v''_3, v''_4\}$) are in the graph of the first (resp., second and third) copy of $\bold{H}_{\textup{OD}}$.}
\label{fig_3}
\vspace{-0.7em}
\end{figure}

\begin{example}\label{ex_3}
Consider an instance of the $(4, 2)$ UAS, $\gamma=3$, in Fig.~\ref{fig_1}, which exists in $\bold{H}_{\textup{OD}}$, and let $M=3$ for $\bold{H}_{\textup{MD}}$. The three copies of the UAS in $\bold{H}_{\textup{MD}}$ are shown in the left panel of Fig.~\ref{fig_2} (degree-$1$ CNs are not shown). We check the following two relocation arrangements:

Arrangement 1: $R \left (\mathcal{E}_{c_5,v_2} \right ) = R \left ( \mathcal{E}_{c_5,v_4} \right ) = 1$, while all the remaining NZ entries of the UAS are not relocated. Here, (\ref{eqn_cyc}) is satisfied for both the dotted blue and the dotted red basic cycles. Thus, the $(4, 2)$ UAS stays active, which is shown in the upper panel of Fig.~\ref{fig_2}.

Arrangement 2: $R \left (\mathcal{E}_{c_3,v_4} \right ) = R \left ( \mathcal{E}_{c_4,v_1} \right ) = 1$, while all the remaining NZ entries of the UAS are not relocated. Here, (\ref{eqn_cyc}) is not satisfied for either basic cycle. Thus, the $(4, 2)$ UAS becomes inactive, which is shown in the lower panel of Fig.~\ref{fig_2}. How the three copies of the $(4, 2)$ UAS result in a $(12, 6)$ object after relocations is demonstrated in Fig.~\ref{fig_3}.
\end{example}

\begin{remark}
The analysis in Sections~\ref{sec_rem} and \ref{sec_save} can be~introduced for NZ circulants, which was the case in \cite{md_res}, rather than NZ entries. However, it is easier for the reader to understand the concepts when NZ entries, or edges, are used.
\end{remark}

\section{Savings in Relocation Options}\label{sec_save}

Targeting UASs instead of the cycles comprising them not only makes the focus in the code design on the more detrimental objects, but also achieves significant savings in the degrees of freedom offered by relocation arrangements. These savings are reflected in performance gains. Here, we demonstrate~these savings. In the following results, fractions are out of all possible relocation arrangements. Let $(x)^+= \max\{x, 0\}$.

Lemma~\ref{lem_cycfr} discusses the relocation arrangements in case the focus is on removing short cycles from the graph of $\bold{H}_{\textup{MD}}$.

\begin{lemma}\label{lem_cycfr}
The fraction of relocation arrangements for an $(a, d_1)$ UAS under which all the basic cycles in a cycle basis $\mathcal{B}_{\textup{c}}$ of the UAS become inactive is given by:
\begin{equation}\label{eqn_fnof}
F_{\textup{nof}} = \left (\frac{M-1}{M} \right )^{\hspace{-0.3em}n_\textup{f}}.
\end{equation}
Moreover, the fraction of relocation arrangements for an $(a, d_1)$ UAS under which all its cycles become inactive is upper-bounded as follows:
\vspace{-0.1em}\begin{equation}\label{eqn_fnoc}
F_{\textup{noc}} \leq \prod_{\delta=1}^{n_\textup{f}} \left ( \frac{M-\delta}{M} \right )^{\hspace{-0.3em}+}.
\end{equation}
\end{lemma}

\begin{IEEEproof}
We can always order the basic cycles in $\mathcal{B}_{\textup{c}}$ of the UAS such that there does not exist a basic cycle sharing all its CNs with previous ones. In this proof, we access the basic cycles one by one according to that order and assign relocations to their edges one by one.

\textbf{Proof of (\ref{eqn_fnof}):} We want to break (\ref{eqn_cyc}) for all the $n_\textup{f}$ basic cycles in $\mathcal{B}_{\textup{c}}$. For the first basic cycle, each of its edges has $M$ different relocation options except the last edge. For that last edge, only $M-1$ relocation options are available since the option that makes (\ref{eqn_cyc}) satisfied is excluded. Assuming that the number of edges in this cycle is $\zeta$, the fraction of relocation arrangements that make this cycle inactive is:
\begin{equation}\label{eqn_fircyc}
\frac{M^{\zeta-1}(M-1)}{M^\zeta} = \frac{M-1}{M}.
\end{equation}
Now, suppose that we are at basic cycle $\delta$, and let $y$ be the number of edges with no relocation assignment after finishing the first $\delta-1$ basic cycles. Note that $y$ has to be greater than $0$ from the order of basic cycles we adopt. Still $y-1$ of those edges have $M$ different relocation options except the last edge, which has only $M-1$ relocation options. Consequently, the fraction of relocation arrangements for the UAS under which all its basic cycles in $\mathcal{B}_{\textup{c}}$ become inactive is:
\begin{equation}
F_{\textup{nof}} = \prod_{\delta=1}^{n_\textup{f}} \left ( \frac{M-1}{M} \right ) = \left (\frac{M-1}{M} \right )^{\hspace{-0.3em}n_\textup{f}}.
\end{equation}

\textbf{Proof of (\ref{eqn_fnoc}):} Here, we adopt the ordering described at the beginning of this proof with one extra condition, which~is each two consecutive basic cycles share at least one degree-$2$ CN (see the proof of Lemma~\ref{lem_ncyc}).

We want to break (\ref{eqn_cyc}) for all the cycles, and we do that via the basic cycles of the UAS. For the first basic cycle, the fraction of relocation arrangements that make this cycle inactive is given by (\ref{eqn_fircyc}). For the second basic cycle, we want not only to make it inactive, but also to make the cycle resulting from adding the vectors of these two basic cycles over GF($2$) inactive. Thus, for the last edge of the second basic cycle, we only have $M-2$ relocation options. The upper bound is satisfied if the lower bound on $n_\textup{c}$ in (\ref{eqn_ncb}) is satisfied. In this case, at basic cycle $\delta$, the number of relocation options we have for the last edge is only $M-\delta$. Consequently,
\begin{align}
F_{\textup{noc}} &\leq \left ( \frac{M-1}{M} \right ) \left ( \frac{M-2}{M} \right ) \cdots \left ( \frac{M-n_{\textup{f}}}{M} \right )^{\hspace{-0.3em}+} \nonumber \\ &= \prod_{\delta=1}^{n_\textup{f}} \left ( \frac{M-\delta}{M} \right )^{\hspace{-0.3em}+},
\end{align}
which completes the proof.
\end{IEEEproof}

Theorem~\ref{thm_objfr} discusses the relocation arrangements in case the focus is on removing UASs from the graph of $\bold{H}_{\textup{MD}}$.

Define $\mathcal{F}_i$ as the set of CNs in basic cycle $i$, and $\mathcal{I}_{i,j} \triangleq \mathcal{F}_{i} \cap \mathcal{F}_{j}$. Moreover,
\begin{equation}\label{eqn_symbs}
\mathcal{I}^{\textup{tot}}_i \triangleq \underset{j}\cup \left ( \mathcal{I}_{i,j} \right ), \textup{ and } \mathcal{D}_i \triangleq \mathcal{F}_i \setminus \mathcal{I}^{\textup{tot}}_i.
\end{equation}
Let $D_i$ be the unordered group comprising the CNs of $\mathcal{D}_i$. Then, we define the following set:
\begin{equation}\label{eqn_l1}
\mathcal{L}_1 \triangleq \{D_i, \forall i \textup{ } | \textup { } \mathcal{D}_i \neq \varnothing \}.
\end{equation}
Let $I_{i,j}$ be the unordered group comprising the CNs of $\mathcal{I}_{i,j}$. Then, we define the following set:
\begin{equation}\label{eqn_l2}
\mathcal{L}_2 \triangleq \{I_{i,j}, \forall i,j \textup{ } | \textup { } \mathcal{I}_{i,j} \neq \varnothing \}.
\end{equation}

\begin{theorem}\label{thm_objfr}
The fraction of relocation arrangements for an $(a, d_1)$ UAS under which the UAS becomes inactive is given by:
\begin{equation}\label{eqn_fnou}
F_{\textup{nou}} = 1 - \frac{1}{M^{n_\textup{f}}}.
\end{equation}
Moreover, the fraction of relocation arrangements for an $(a, d_1)$ UAS under which the $M$ copies of the UAS result in at least $M$ $(a, d_1+2\beta)$ objects, with $\beta > 1$, is given by:
\begin{equation}\label{eqn_fnot}
F_{\textup{not}} = 1 - \frac{1}{M^{n_\textup{f}}} - \left [ \vert \mathcal{L}_1 \vert + \vert \mathcal{L}_2 \vert \right ] \frac{(M-1)}{M^{n_\textup{f}}}.
\end{equation}
\end{theorem}

\begin{IEEEproof}
We order the basic cycles in a cycle basis $\mathcal{B}_{\textup{c}}$ of the UAS as done in the proof of Lemma~\ref{lem_cycfr}. We also access the basic cycles and assign relocations to their edges one by one according to that order.

\textbf{Proof of (\ref{eqn_fnou}):}
First, we find the fraction of relocation~arrangements under which the UAS stays active. Thus, and from Theorem~\ref{thm_cond}, we want to satisfy (\ref{eqn_cyc}) for all the $n_\textup{f}$ basic cycles in $\mathcal{B}_{\textup{c}}$ to make them active. For the first basic cycle, each of its edges has $M$ different relocation options except the last edge. For that last edge, only $1$ relocation option is available to satisfy (\ref{eqn_cyc}). Assuming that the number of edges in this cycle is $\zeta$, the fraction of relocation arrangements that make this cycle active is:
\begin{equation}\label{eqn_fircyc2}
\frac{M^{\zeta-1}(1)}{M^\zeta} = \frac{1}{M}.
\end{equation}
Now, suppose that we are at basic cycle $\delta$, and let $y$ be the number of edges with no relocation assignment after finishing the first $\delta-1$ basic cycles. Still $y-1$ of those edges have $M$ different relocation options except the last edge, which has only $1$ relocation option. Consequently, the fraction of relocation arrangements for the UAS under which all its basic cycles in $\mathcal{B}_{\textup{c}}$ stay active is:
\vspace{-0.1em}\begin{equation}
F_0 = \prod_{\delta=1}^{n_\textup{f}} \left ( \frac{1}{M} \right ) = \frac{1}{M^{n_\textup{f}}}.
\end{equation}
From the definition of $F_{\textup{nou}}$, we infer:
\begin{equation}\label{eqn_fnoupr}
F_{\textup{nou}} = 1 - F_0 = 1 - \frac{1}{M^{n_\textup{f}}}.
\end{equation}

\textbf{Proof of (\ref{eqn_fnot}):} We find the fraction of relocation~arrangements under which the $M$ copies of the UAS result in at least $M$ $(a, d_1+2)$ objects. Observe that an $(a, d_1+2)$ object is the result of disconnecting exactly one degree-$2$ CN from the $(a, d_1)$ UAS. In order for this to happen, one of following two scenarios has to happen.

The first scenario is that only one basic cycle becomes inactive after relocations, while the remaining basic cycles stay active. Moreover, this basic cycle must have at least one CN that is not shared with any other basic cycles. The last edge in the basic cycle that is to be inactive has $M-1$ relocation options. The last edge in each of the remaining $n_{\textup{f}}-1$ basic cycles has only $1$ relocation option. Consequently, and using the definition of $\mathcal{L}_1$ in (\ref{eqn_l1}), the fraction of relocation arrangements satisfying the first scenario is:
\begin{equation}\label{eqn_add1}
\vert \mathcal{L}_1 \vert \left ( \frac{M-1}{M} \right ) \prod_{\delta=1}^{n_\textup{f}-1} \left ( \frac{1}{M} \right ) = \vert \mathcal{L}_1 \vert \frac{(M-1)}{M^{n_\textup{f}}}.
\end{equation}

The second scenario is that only two basic cycles become inactive after relocations, while the remaining basic cycles stay active. Moreover, these two basic cycles must have at least one CN that is shared between them, and the cycle resulting from adding the vectors of these two basic cycles over GF($2$) must stay active. The last edge in the first basic cycle that is to be inactive has $M-1$ relocation options. The last edge in the second basic cycle that is to be inactive has only $1$ relocation option (that makes it inactive but keeps the cycle resulting from adding the vectors of the two basic cycles active). The last edge in each of the remaining $n_{\textup{f}}-2$ basic cycles has only $1$ relocation option. Consequently, and using the definition of $\mathcal{L}_2$ in (\ref{eqn_l2}), the fraction of relocation arrangements satisfying the second scenario is:
\begin{equation}\label{eqn_add2}
\vert \mathcal{L}_2 \vert \left ( \frac{M-1}{M} \right ) \left ( \frac{1}{M} \right ) \prod_{\delta=1}^{n_\textup{f}-2} \left ( \frac{1}{M} \right ) = \vert \mathcal{L}_2 \vert \frac{(M-1)}{M^{n_\textup{f}}},
\end{equation}
where using $\vert \mathcal{L}_2 \vert$ filters out repeated groups of CNs. Note that similar scenarios dealing with more than two basic cycles will result in groups of CNs already in $\mathcal{L}_2$. Thus, the fraction of relocation arrangements under which the $M$ copies of the UAS result in at least $M$ $(a, d_1+2)$ objects is obtained by adding (\ref{eqn_add1}) and (\ref{eqn_add2}):
\begin{equation}
F_1 = \left [ \vert \mathcal{L}_1 \vert + \vert \mathcal{L}_2 \vert \right ] \frac{(M-1)}{M^{n_\textup{f}}}.
\end{equation}
From the definition of $F_{\textup{not}}$, we infer:
\begin{align}\label{eqn_fnotpr}
F_{\textup{not}} &= 1 - F_0 - F_1 \nonumber \\ &= 1 - \frac{1}{M^{n_\textup{f}}} - \left [ \vert \mathcal{L}_1 \vert + \vert \mathcal{L}_2 \vert \right ] \frac{(M-1)}{M^{n_\textup{f}}},
\end{align}
which completes the proof.
\end{IEEEproof}

On the level of an object, the percentage saving in relocation arrangements achieved by focusing on the UAS instead of focusing on all its cycles is given by:
\begin{align}\label{eqn_save1}
S_1 &= \left [ F_{\textup{nou}} - \textup{bound}(F_{\textup{noc}}) \right ] \cdot 100\% \nonumber \\ &= \left [ 1 - \frac{1}{M^{n_\textup{f}}} - \prod_{\delta=1}^{n_\textup{f}} \left ( \frac{M-\delta}{M} \right )^{\hspace{-0.3em}+} \right ] \cdot 100\%.
\end{align}

If $d_1$ is in $\{0,1\}$, and all $(a, d_1+2)$ configurations are not desirable when focusing on the UAS, while only making all the basic cycles in $\mathcal{B}_{\textup{c}}$ inactive is enough when focusing on cycles, a stricter formula for the percentage saving in relocation arrangements should be used:
\begin{align}\label{eqn_save2}
S_2 &= \left [ F_{\textup{not}} - F_{\textup{nof}} \right ] \cdot 100\% \nonumber \\ &= \left [ 1 - \frac{1}{M^{n_\textup{f}}} - \left [ \vert \mathcal{L}_1 \vert + \vert \mathcal{L}_2 \vert \right ] \frac{(M-1)}{M^{n_\textup{f}}} - \left (\frac{M-1}{M} \right )^{\hspace{-0.3em}n_\textup{f}} \right ] \nonumber \\ &\hspace{+1.0em} \cdot 100\%.
\end{align}
In fact, $S_1$ (resp., $S_2$ if applicable) can be viewed as the ceiling (resp., floor) of the percentage saving in relocation options.

\begin{example}\label{ex_4}
Consider the $(4, 2)$ UAS, $\gamma=3$, in Fig~\ref{fig_1}. Let $M=5$. From Example~\ref{ex_1}, $n_{\textup{f}}=2$. Thus, from (\ref{eqn_fnof}) and (\ref{eqn_fnoc}),
\begin{align}
F_{\textup{nof}} &= \left (\frac{4}{5} \right )^{\hspace{-0.3em}2} = \frac{16}{25}, \nonumber \\ F_{\textup{noc}} &\leq \prod_{\delta=1}^{2} \left ( \frac{5-\delta}{5} \right )^{\hspace{-0.3em}+} = \left ( \frac{4}{5} \right ) \left ( \frac{3}{5} \right ) = \frac{12}{25}. \nonumber
\end{align}

The two basic cycles here have $\mathcal{F}_1 = \{c_1, c_4, c_5 \}$ and $\mathcal{F}_2 = \{c_2, c_3, c_5 \}$.
Consequently, we get $\mathcal{I}_{1,2} = \{c_5\}$, yielding $\mathcal{I}^{\textup{tot}}_1 = \mathcal{I}^{\textup{tot}}_2 = \{c_5 \}$. From (\ref{eqn_symbs}), $\mathcal{D}_1 = \{ c_1, c_4 \}$ and $\mathcal{D}_2 = \{ c_2, c_3 \}$. Thus, from (\ref{eqn_l1}) and (\ref{eqn_l2}), we get:
\begin{equation}
\mathcal{L}_1 = \{ (c_1, c_4), (c_2, c_3) \}, \textup{ }
\mathcal{L}_2 = \{ (c_5) \}. \nonumber
\end{equation}
From (\ref{eqn_fnou}) and (\ref{eqn_fnot}),
\begin{align}
F_{\textup{nou}} &= 1 - \frac{1}{5^2} = \frac{24}{25}, \nonumber \\ F_{\textup{not}} &= 1 - \frac{1}{5^2} - [ 2+1 ] \frac{4}{5^2} = \frac{12}{25}. \nonumber
\end{align}

Now, we are ready to calculate the saving in relocation arrangements from (\ref{eqn_save1}) as follows:
\begin{align}
S_1 = \left [ \frac{24}{25} - \frac{12}{25} \right ] \cdot 100\% = 48\%, \nonumber
\end{align}
which is a significant saving.
\end{example}

\begin{example}\label{ex_5}
Consider the $(4, 4)$ UAS, $\gamma=4$, in Fig~\ref{fig_1}. Let $M=5$. From Example~\ref{ex_2}, $n_{\textup{f}}=3$. Thus, from (\ref{eqn_fnof}) and (\ref{eqn_fnoc}),
\begin{align}
F_{\textup{nof}} &= \left (\frac{4}{5} \right )^{\hspace{-0.3em}3} = \frac{64}{125}, \nonumber \\ F_{\textup{noc}} &\leq \prod_{\delta=1}^{3} \left ( \frac{5-\delta}{5} \right )^{\hspace{-0.3em}+} = \left ( \frac{4}{5} \right ) \left ( \frac{3}{5} \right ) \left ( \frac{2}{5} \right ) = \frac{24}{125}. \nonumber
\end{align}

The three basic cycles here have $\mathcal{F}_1 = \{c_1, c_4, c_5 \}$, $\mathcal{F}_2 = \{c_2, c_3, c_5 \}$, and $\mathcal{F}_3 = \{c_3, c_4, c_6 \}$.
Consequently, we get $\mathcal{I}_{1,2} = \{c_5\}$, $\mathcal{I}_{1,3} = \{c_4\}$, and $\mathcal{I}_{2,3} = \{c_3\}$, yielding $\mathcal{I}^{\textup{tot}}_1 = \{c_4, c_5 \}$,  $\mathcal{I}^{\textup{tot}}_2 = \{c_3, c_5 \}$, and  $\mathcal{I}^{\textup{tot}}_3 = \{c_3, c_4 \}$. From (\ref{eqn_symbs}), $\mathcal{D}_1 = \{ c_1 \}$, $\mathcal{D}_2 = \{ c_2 \}$, and $\mathcal{D}_3 = \{ c_6 \}$. Thus, from (\ref{eqn_l1}) and  (\ref{eqn_l2}), we get:
\begin{equation}
\mathcal{L}_1 = \{ (c_1), (c_2), (c_6) \}, \textup{ }
\mathcal{L}_2 = \{ (c_5), (c_4), (c_3) \}. \nonumber
\end{equation}
From (\ref{eqn_fnou}) and (\ref{eqn_fnot}),
\begin{align}
F_{\textup{nou}} &= 1 - \frac{1}{5^3} = \frac{124}{125}, \nonumber \\ F_{\textup{not}} &= 1 - \frac{1}{5^3} - [ 3+3 ] \frac{4}{5^3} = \frac{100}{125}. \nonumber
\end{align}

Now, we are ready to calculate the saving in relocation arrangements from (\ref{eqn_save1}) as follows:
\begin{align}
S_1 = \left [ \frac{124}{125} - \frac{24}{125} \right ] \cdot 100\% = 80\%, \nonumber
\end{align}
which is a significant saving.

Observe that the same analysis is applicable for the $(4, 0)$ UAS, $\gamma=3$, where all degree-$1$ CNs are eliminated. In this case, and using (\ref{eqn_save2}), $ S_2 = \left [ \frac{100}{125} - \frac{64}{125} \right ] \cdot 100\% \approx 29\%$ also becomes useful.
\end{example}

Now, we briefly introduce a special case of interest.

\begin{definition}\label{def_regst}
Let $a_{\textup{min}}$ be the minimum UAS size in the OD code. An $(a, d_1)$ UAS, $a < Ma_{\textup{min}}$, is said to be non-regenerable if it cannot be produced from $(a, d_1^-)$ UASs, $\psi^- < \psi$, under any relocation arrangement. Furthermore, an $(a, d_1)$ UAS is said to be stand-alone if an instance of this UAS cannot share any cycles with another instance of it.
\end{definition}

For example, $(a, 0)$ UASs are non-regenerable and stand-alone. Additionally, UASs with $d_2 = \binom{a}{2}$ are non-regenerable.

For non-regenerable, stand-alone UASs, the savings in relocation arrangements given in (\ref{eqn_save1}) and (\ref{eqn_save2}) can be generalized over the entire graph of the MD code. More intriguingly, under random relocations, the average number of instances of an $(a, d_1)$ non-regenerable, stand-alone UAS in the graph of the MD code is given by:
\begin{equation}\label{eqn_amd}
\overline{A}_{\textup{MD}} = A_{\textup{OD}} F_0 M,
\end{equation}
where $A_{\textup{OD}}$ is the number of instances in the OD code, and $F_0 = \frac{1}{M^{n_\textup{f}}}$. The average for regenerable, stand-alone $(a, 2)$ UASs can also be found. These averages give the code designer an initial idea about the optimization effort to be exerted to design the MD code. Thus, deriving these averages for any $(a, d_1)$ UAS is an interesting research problem.

\section{Algorithm and Experimental Results}\label{sec_exp}

We are now ready to introduce the algorithm using which, we design our high performance MD codes. Guided by the previously illustrated theoretical results, Algorithm~\ref{alg_md} minimizes the number of instances of a specific $(a, d_1)$ UAS, $a < Ma_{\textup{min}}$, in the graph of the MD code via relocations. This specific $(a, d_1)$ UAS/AS can either be the most dominant object in the error profile of the OD code or a common substructure that exists in the most dominant UASs in the OD code. Determining this $(a, d_1)$ UAS depends on both the channel of interest \cite{ahh_tit, ahh_nboo} and the OD code being used.

Because of their faster encoding and decoding, we focus on graph-based codes that are circulant-based in this section. Since operating on circulants is significantly faster than operating on entries, Algorithm~\ref{alg_md} relocates NZ circulants, not NZ entries (see also \cite{md_res}). The algorithm can be easily changed to relocate NZ entries for codes that are not structured.

We say that an $(a, d_1)$ UAS instance \textbf{involves} a circulant if the instance has at least one NZ entry corresponding to an edge adjacent to a degree-$2$ CN inside the circulant. Moreover, the set of relocation decisions is $\mathcal{X}=\{0, 1, \dots, M-1\}$. The value $\xi \in \mathcal{X}$, $\xi > 0$ (resp., $\xi = 0$), refers to the decision ``relocate to $\bold{X}_\xi$'' (resp., ``no relocation'').

Note that in Step~3 of Algorithm~\ref{alg_md}, if the OD code is SC, its repetitive nature should be exploited in order to reduce the processing time. Note also that Step~16 of Algorithm~\ref{alg_md} aims to balance the number of NZ circulants (similar sparsity levels) across all auxiliary matrices in addition to its main objective, which is removing $(a, d_1)$ UAS instances.

\begin{algorithm}[H]
\caption{Designing High Performance MD Codes}
\begin{algorithmic}[1]
\State \textbf{Inputs:} $\bold{H}_{\textup{OD}}$, $M$, and the $(a, d_1)$ UAS configuration.
\State Initially, set $\bold{X}_1 = \bold{X}_2 = \cdots = \bold{X}_{M-1} = \bold{0}$, $\bold{H}'_{\textup{OD}} = \bold{H}_{\textup{OD}}$, and $R(\mathcal{E}_{i,j}) = 0$, $\forall i,j$.
\State Locate all instances of the $(a, d_1)$ UAS in the graph of $\bold{H}_{\textup{OD}}$.
\State Mark all the instances located in Step~3 as active.
\State Determine the number of active $(a, d_1)$ UAS instances involving each NZ circulant in $\bold{H}_{\textup{OD}}$.
\State Select the circulant $\mathcal{C}$ with the maximum number from Step~5 s.t. $R(\mathcal{E}_{i_t,j_t}) = 0$, where $\mathcal{E}_{i_t,j_t}$ is an NZ entry in~$\mathcal{C}$.
\State Whether they are active or not, specify all $(a, d_1)$ UAS instances in $\bold{H}_{\textup{OD}}$ involving $\mathcal{C}$.
\State \textbf{for} each of the instances from Step~7 \textbf{do}
\State \hspace{2ex} Specify a cycle basis $\mathcal{B}_{\textup{c}}$ of the instance.
\State \hspace{2ex} The instance votes for the subset of decisions in $\mathcal{X}$ that make at least one of its basic cycles in $\mathcal{B}_{\textup{c}}$ inactive.
\State \textbf{end for}
\State Tally the votes, and find the subset $\mathcal{W}$ of NZ decisions in $\mathcal{X}$ with the highest number of votes.
\State \textbf{if} $\mathcal{W} = \varnothing$ \textbf{then} \textit{(no relocation)}
\State \hspace{2ex} Go to Step~22.
\State \textbf{end if}
\State Relocate $\mathcal{C}$ to the auxiliary matrix $\bold{X}_{\xi_r}$, $\xi_r \in \mathcal{W}$, with the least number of NZ circulants.
\State Set $R(\mathcal{E}_{i,j}) = \xi_r$ for all NZ entries in $\mathcal{C}$. 
\State Update the list of active/inactive $(a, d_1)$ instances based on their basic cycles and Theorem~\ref{thm_cond}.
\State \textbf{if} the number of active $(a, d_1)$ instances is $> 0$ \textbf{then}
\State \hspace{2ex} Go to Step~5.
\State \textbf{end if}
\State Construct $\bold{H}_{\textup{MD}}$ according to (\ref{eqn_hmd}).
\State \textbf{Output:} The parity-check matrix of the MD code, $\bold{H}_{\textup{MD}}$.
\end{algorithmic}
\label{alg_md}
\end{algorithm}

Here, we assume that if the UAS entered at Step~1 of the algorithm is an $(a, d_1)$ UAS, then all possible $(a, d_1^-)$ UASs do not exist in the OD code. Thus, Algorithm~\ref{alg_md} works for regenerable as well as non-regenerable UASs.

Extending Algorithm~\ref{alg_md} to operate on multiple detrimental configurations is possible. In this case, different UASs should be ordered according to the values of $a$ and $d_1$ from the smallest to the largest, and the algorithm should operate on them successively. However, this extension is associated with a challenge; that is, objects having $a^-$ VNs or/and $d_2^-$ degree-$2$ CNs in the OD code may result in $(a, d_1)$ UASs in the MD code after relocations. Resolving this challenge to implement the extension is another interesting problem. Observe that in OD codes with no cycles of length $4$, and if $a < 6$, $(a, d_1)$ UASs cannot be generated from smaller objects, i.e., objects having $a^-$ VNs or/and $d_2^-$ degree-$2$ CNs.

\begin{remark}
The concept of basic cycles can be used to determine the conditions under which cycles of certain lengths in the OD code result in a bigger cycle in the MD code after relocations. Thus, this concept can also be used to determine whether an $(a, d_1)$ UAS can be generated from smaller objects under certain relocations.
\end{remark}

\begin{remark}
In the construction procedure of $\bold{H}_{\textup{MD}}$, circulants are relocated from the copies of $\bold{H}_{\textup{OD}}$ to the auxiliary matrices in the exact same positions. Thus, the structure of all submatrices in $\bold{H}_{\textup{MD}}$ resembles the structure of $\bold{H}_{\textup{OD}}$. Decoding algorithms can be derived to exploit this property, significantly reducing the decoding latency of MD codes.
\end{remark}

Next, we discuss the experimental results. The Flash channel used in this section is a practical, asymmetric Flash channel, which is the normal-Laplace mixture (NLM) Flash channel \cite{mit_nl}. In the NLM channel, the threshold voltage distribution of sub-$20$nm multi-level cell (MLC) Flash memories is carefully modeled. The four levels are modeled as different NLM distributions, incorporating several sources of error due to wear-out effects, e.g., programming errors, thereby resulting in significant asymmetry. Furthermore, the authors provided accurate fitting results of their model for program/erase (P/E) cycles up to $10$ times the manufacturer's endurance specification (up to $30000$ P/E cycles). We implemented the NLM channel based on the parameters described in \cite{mit_nl}. Here, we use $3$ reads, and the sector size is $512$ bytes. For decoding, we use a fast Fourier transform based $q$-ary sum-product algorithm (FFT-QSPA) LDPC decoder (see also \cite{ahh_tit}).

We use three OD codes in this section. The SC codes are designed according to \cite{ahh_nboo}, which provides a method to design high performance SC codes particularly for Flash systems. This method is based on the optimal overlap, circulant power optimizer (OO-CPO) approach. The block code is designed according to \cite[Section~VI]{ahh_tit}. OD~Code~1 is an SC code defined over GF($4$), which has $\gamma=3$, maximum row weight $=19$, circulant size $=19$, memory $=1$, and coupling length $=7$. Thus, OD~Code~1 has block length $=5054$ bits and rate $\approx 0.82$. OD~Code~2 is a block code defined over GF($2$), which has $\gamma=4$, row weight $=40$, and circulant size $=53$. OD~Code~2 has block length $=4240$ bits and rate $\approx 0.90$. OD~Code~1 and OD~Codes~2 are the underlying codes of our MD codes. OD~Code~3 is an SC code that is designed exactly as OD~Code~1, except for that OD~Code~3 has coupling length $=21$ instead of $7$ (three times as long as OD~Code~1). Thus, OD~Code~3 has block length $=15162$ bits and rate $\approx 0.83$.

From our simulations, the error profile in the error floor region of OD~Code~1 when simulated over the NLM channel is dominated by the $(4, 2)$ non-binary AS. In fact, this is a general AS of type two (GAST) according to \cite{ahh_tit}, but we abbreviate the notation here for simplicity. Moreover, the error profile in the error floor region of OD~Code~2 when simulated over the AWGN channel is dominated by the $(4, 4)$ and the $(6, 2)$ UASs. The overwhelming majority of the $(6, 2)$ UAS instances found in the error profile of OD~Code~2 simulated over the AWGN channel have the same configuration, which has the $(4, 4)$ UAS as a substructure. Note that for a binary code, e.g., OD~Code~2, a UAS is an AS. Note also that OD~Code~1 and OD~Code~2 are the underlying OD codes of the MD codes used in this section.

\begin{remark}\label{rmk_adv}
The objects of interest in other codes and over other channels can be more sophisticated, e.g., the $(6, 0)$ and the $(8, 0)$ UASs, $\gamma = 3$, in addition to the $(6, 6)$ and the $(8, 2)$ UASs, $\gamma = 4$. See \cite{ahh_tit} for more details.
\end{remark}

As for the MD codes, MD~Code~1 is designed for practical Flash channels, while MD~Code~2 is designed for AWGN channels. According to the analysis above, MD~Code~1, with $M=3$, is designed from OD~Code~1 using Algorithm~\ref{alg_md} as follows. Algorithm~\ref{alg_md} is used to remove as many $(4, 2)$ UAS instances as possible in the MD code via relocations since the $(4, 2)$ UAS is the unlabeled configuration of the most dominant AS over the NLM channel. Furthermore, MD~Code~2, with $M=3$, is designed from OD~Code~2 using Algorithm~\ref{alg_md} as follows. Algorithm~\ref{alg_md} is used to remove as many $(4, 4)$ UAS instances as possible in the MD code via relocations since the $(4, 4)$ UAS is the common substructure of interest over the AWGN channel. MD~Code~1 has block length $=15162$ bits and rate $\approx 0.82$, which is similar to OD~Code~3 (the long OD SC code described above). MD~Code~2 has block length $=12720$ bits and rate $\approx 0.90$. No specific optimization is performed to the edge weights of non-binary codes.

\begin{table}
\caption{Effect of carefully chosen MD relocations on the number of $(4, 2)$ UAS instances, $\gamma=3$.}
\vspace{-0.5em}
\centering
\scalebox{1.00}
{
\begin{tabular}{|c|c|}
\hline
\makecell{MD coupling \\ technique} & \makecell{Number of $(4, 2)$ \\ UAS instances} \\
\hline
No MD coupling & $4218$ \\
\hline
Algorithm~\ref{alg_md} & $0$ \\
\hline
\end{tabular}}
\label{table_1}
\vspace{-0.5em}
\end{table}

\begin{table}
\caption{Effect of carefully chosen MD relocations on the number of $(4, 4)$ UAS instances, $\gamma=4$.}
\vspace{-0.5em}
\centering
\scalebox{1.00}
{
\begin{tabular}{|c|c|}
\hline
\makecell{MD coupling \\ technique} & \makecell{Number of $(4, 4)$ \\ UAS instances} \\
\hline
No MD coupling & $3392$ \\
\hline
Algorithm~\ref{alg_md} & $0$ \\
\hline
\end{tabular}}
\label{table_2}
\vspace{-0.7em}
\end{table}

Table~\ref{table_1} and Table~\ref{table_2} demonstrate the reduction in the number of UASs achieved by Algorithm~\ref{alg_md}. The no-MD-coupling case refers to the case when $\bold{H}_{\textup{MD}}$ is constructed by putting three copies of $\bold{H}_{\textup{OD}}$ in the block diagonal and zeros elsewhere. Table~\ref{table_1} shows that, and with only about $7\%$ of the circulants relocated out of the OD copies to construct MD~Code~1, Algorithm~\ref{alg_md} removes all the $(4, 2)$ UAS instances. Additionally, Table~\ref{table_2} shows that, and with only about $4.5\%$ of the circulants relocated out of the OD copies to construct MD~Code~2, Algorithm~\ref{alg_md} removes all the $(4, 4)$ UAS instances. These relatively small percentages of relocated circulants exemplify the savings in relocation arrangements (see Section~\ref{sec_save}) in the MD code design, making it possible to relocate more circulants in order to remove other detrimental objects.

In this section, RBER is the raw bit error rate, which is the number of raw, i.e., uncoded, data bits in error divided by the total number of raw data bits read \cite{ahh_tit}. UBER is the uncorrectable bit error rate, which is a metric for the fraction of bits in error out of all bits read after the error correction is applied. Here, the formulation of UBER is the frame error rate (FER) divided by the sector size in bits \cite{ahh_tit}.

\begin{figure}
\vspace{-0.3em}
\center
\includegraphics[trim={0.3in 0.0in 0.9in 0.3in}, width=3.4in]{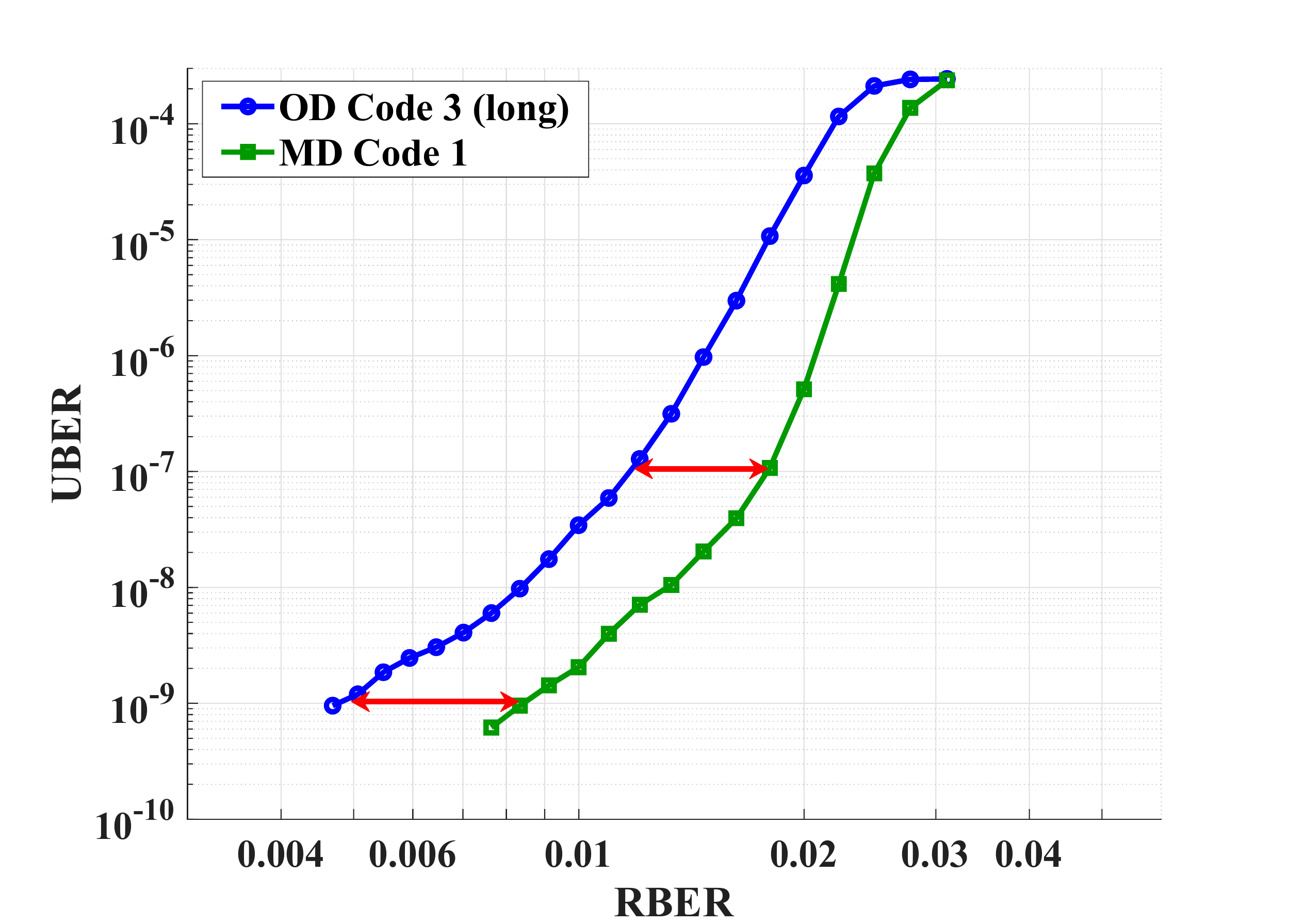}
\vspace{-0.5em}
\caption{UBER versus RBER curves over the NLM Flash channel for OD and MD codes of similar parameters.}
\label{fig_4}
\vspace{-1.0em}
\end{figure}

Fig.~\ref{fig_4} demonstrates the performance gains achieved by an MD code constructed using Algorithm~\ref{alg_md}, which is MD~Code~1, compared with an OD code of similar length and rate, which is OD~Code~3, over the practical NLM Flash channel. In particular, at UBER $\approx 10^{-7}$ in the waterfall region, the RBER gain of MD~Code~1 marked in red translates to a gain of about $1200$ P/E cycles. Moreover, at UBER $\approx 10^{-9}$ in the error floor region, the RBER gain of MD~Code~1 marked in red translates to a gain of about $1800$ P/E cycles. In addition to the waterfall slope and the error floor slope/level, even the threshold of MD~Code~1 is indeed better than that of OD~Code~3. These gains in the number of P/E cycles are associated with an increase in the lifetime of the Flash device.

\begin{remark}
The error floor performance of both non-binary codes having their performance curves in Fig.~\ref{fig_4}, which are OD~Code~3 and MD~Code~1, can be improved using the weight consistency matrix (WCM) framework described in \cite{ahh_tit}.
\end{remark}

\begin{remark}\label{rmk_sims}
While we focus here on practical Flash channels in the simulations, performance gains are also achievable via the proposed technique on other channels.
\end{remark}

\section{Conclusion}\label{sec_conc}

We introduced necessary and sufficient conditions for a UAS to stay active or become inactive, i.e., be removed, after a relocation arrangement. We derived the savings in relocation options achieved by focusing on UASs instead of cycles in the MD code design procedure. Examples demonstrating the significance of these savings were introduced for famous UAS configurations. We presented an algorithm to design high performance MD codes by removing detrimental UASs via relocations. Using this algorithm, codes free of specific UASs were designed and simulated. Gains of up to about $1800$ P/E cycles were achieved via our MD codes compared with OD codes of similar parameters over a practical Flash channel.

\section*{Acknowledgment}\label{sec_ack}

This research was supported in part by NSF under grant CCF 1717602.



\vspace{-0.3em}
\begin{thebibliography}{13}

\balance

\bibitem{mit_nl}
T. Parnell, N. Papandreou, T. Mittelholzer, and H. Pozidis, ``Modelling of the threshold voltage distributions of sub-20nm NAND flash memory,'' in \emph{Proc. IEEE Global Commun. Conf. (GLOBECOM)}, Austin, TX, USA, Dec. 2014, pp. 2351--2356.

\bibitem{ahh_tit}
A. Hareedy, C. Lanka, N. Guo, and L. Dolecek, ``A combinatorial methodology for optimizing non-binary graph-based codes: theoretical analysis and applications in data storage,'' \emph{IEEE Trans. Inf. Theory}, vol. 65, no. 4, pp. 2128--2154, Apr. 2019.

\bibitem{shafa}
S. Srinivasa, Y. Chen, and S. Dahandeh, ``A communication-theoretic framework for 2-DMR channel modeling: performance evaluation of coding and signal processing methods,'' \emph{IEEE Trans. Magn.}, vol. 50, no. 3, pp. 6--12, Mar. 2014.

\bibitem{chen_2d}
P. Chen, C. Kui, L. Kong, Z. Chen, M. Zhang, ``Non-binary protograph-based LDPC codes for 2-D-ISI magnetic recording channels,'' \emph{IEEE Trans. Magn.}, vol. 53, no. 11, Nov. 2017, Art. no. 8108905.

\bibitem{truhach}
D. Truhachev, D. G. M. Mitchell, M. Lentmaier, and D. J. Costello, ``New codes on graphs constructed by connecting spatially coupled chains,'' in \emph{Proc. Inf. Theory and App. Workshop (ITA)}, Feb. 2012, pp. 392--397.

\bibitem{ohashi}
R. Ohashi, K. Kasai, and K. Takeuchi, ``Multi-dimensional spatially-coupled codes,'' in \emph{Proc. IEEE Int. Symp. Inf. Theory (ISIT)}, Jul. 2013, pp. 2448--2452.

\bibitem{schmalen}
L. Schmalen and K. Mahdaviani, ``Laterally connected spatially coupled code chains for transmission over unstable parallel channels,'' in \emph{Proc. Int. Symp. Turbo Codes Iterative Inf. Processing (ISTC)}, Aug. 2014, pp. 77--81.

\bibitem{liu}
Y. Liu, Y. Li, and Y. Chi, ``Spatially coupled {LDPC} codes constructed by parallelly connecting multiple chains,'' \emph{IEEE Commun. Letters}, vol.~19, no.~9, pp. 1472--1475, Sep. 2015.

\bibitem{md_res}
H. Esfahanizadeh, A. Hareedy, and L. Dolecek, ``Multi-dimensional spatially-coupled code design
through informed relocation of circulants,'' in \emph{Proc. 56th Annual Allerton Conf. Commun., Control, and Computing}, Monticello, IL, USA, Oct. 2018, pp. 695--701.

\bibitem{lara_as}
L. Dolecek, Z. Zhang, V. Anantharam, M. Wainwright, and B. Nikolic, ``Analysis of absorbing sets and fully absorbing sets of array-based LDPC codes,'' \emph{IEEE Trans. Inf. Theory}, vol. 56, no. 1, pp. 181--201, Jan. 2010.

\bibitem{fos_lift}
M. P. C. Fossorier, ``Quasi-cyclic low-density parity-check codes from circulant permutation matrices,'' \emph{IEEE Trans. Inf. Theory}, vol. 50, no. 8, pp. 1788--1793, Aug. 2004.

\bibitem{behzad_elem}
B. Amiri, J. Kliewer, and L. Dolecek, ``Analysis and enumeration of absorbing sets for non-binary graph-based codes,'' \emph{IEEE Trans. Commun.}, vol. 62, no. 2, pp. 398--409, Feb. 2014.

\bibitem{ahh_nboo}
A. Hareedy, H. Esfahanizadeh, and L. Dolecek, ``High performance non-binary spatially-coupled codes for Flash memories,'' in \emph{Proc. IEEE Inf. Theory Workshop (ITW)}, Kaohsiung, Taiwan, Nov. 2017, pp. 229--233.

\end{thebibliography}
\end{document}